\documentclass[sigconf,screen]{acmart}
\usepackage{amsmath}
\usepackage{xspace}
\usepackage{tcolorbox}
\usepackage{listings}
\usepackage{xcolor}
\usepackage{titlesec}
\usepackage{enumitem}
\usepackage[ruled,vlined]{algorithm2e}
\usepackage{algpseudocode}
\usepackage{adjustbox}
\usepackage{booktabs}
\usepackage{hyperref}
\usepackage[normalem]{ulem}
\usepackage{colortbl}
\usepackage{multirow}
\usepackage{hhline}
\usepackage{pdflscape}
\usepackage{pifont}
\usepackage{latexsym}
\usepackage{stmaryrd}
\usepackage{balance}
\usepackage{caption}
\usepackage{wasysym}
\usepackage[flushmargin,bottom]{footmisc}
\usepackage{comment}
\usepackage{tikz}

\newtcolorbox{conclusionbox}{colback=gray!8,colframe=black,width=\linewidth,arc=0.6mm, boxrule=0.4pt, left=1mm,right=1mm,top=1mm,bottom=1mm}

\newcommand{\cmark}{\LEFTcircle}%
\newcommand{\cxmark}{\RIGHTcircle}%
\newcommand{\xmark}{\Circle}

\newcommand{\authnote}[2]{{\bf \textcolor{blue}{#1}: \em \textcolor{red}{#2}}}
\newcommand{\yc}[1]{\authnote{YC}{#1}}

\newcommand{\query}[1]
{\small{\textsf{\textcolor{teal}{#1}}}\normalsize}

\newcommand{\tablequery}[1]{\small{\textsf{\textcolor{black}{#1}}}\normalsize}

\newcommand{\name}{\textsf{SQLxDiff}\xspace}

\newcommand{\totalbug}{57\xspace}
\newcommand{\logicbug}{17\xspace}
\newcommand{\abortbug}{18\xspace}
\newcommand{\errorbug}{40\xspace}
\newcommand{\fixbug}{50\xspace}
\newcommand{\confbug}{5\xspace}

\newcommand{\approachabbr}{\textsc{Xxx}\xspace}

\newcommand*\circled[1]{\tikz[baseline=(char.base)]{\node[shape=circle,fill,inner sep=0.7pt] (char) {\textcolor{white}{#1}};}}

\newcommand{\textie}{\textit{i.e.,}\xspace}
\newcommand{\texteg}{\textit{e.g.,}\xspace}

\usepackage{relsize}

\newcommand{\add}[1]{}
\renewcommand{\add}[1]{{\color{red} {#1}}}
\newcommand{\del}[1]{}
\renewcommand{\del}[1]{{\color{blue} \sout{#1}}}

\titlespacing*{\section}{0pt}{0.5\baselineskip}{0.5\baselineskip}
\titlespacing*{\subsection}{0pt}{0.4\baselineskip}{0.4\baselineskip}
\titlespacing*{\subsubsection}{0pt}{0.3\baselineskip}{0.3\baselineskip}

\definecolor{keyword}{HTML}{2771a3}
\definecolor{pattern}{HTML}{b53c2f}
\definecolor{string}{HTML}{be681c}
\definecolor{relation}{HTML}{7e4894}
\definecolor{variable}{HTML}{107762}
\definecolor{comment}{HTML}{8d9094}

\lstdefinelanguage{cypher}
{
	morekeywords={
		CALL, MATCH, OPTIONAL, WHERE, NOT, AND, OR, XOR, RETURN, DISTINCT, ORDER, BY, ASC, ASCENDING, DESC, DESCENDING, UNWIND, AS, UNION, WITH, ALL, CREATE, DELETE, DETACH, REMOVE, SET, MERGE, SET, SKIP, LIMIT, IN, CASE, WHEN, THEN, ELSE, END,
		INDEX, DROP, UNIQUE, CONSTRAINT, EXPLAIN, PROFILE, START, IS
	},
	comment = [l]{//},
}

\definecolor{dkgreen}{rgb}{0,.6,0}
\definecolor{dkblue}{rgb}{0,0,.6}
\definecolor{dkyellow}{cmyk}{0,0,.8,.3}

\setcopyright{None}
\acmConference[Conf '25]{2025 The International Symposium on Software Testing and Analysis (ISSTA)}{June 25--28, 2025}{Trondheim, Norway}
\acmYear{2025}


\begin{document}

\date{}

\title{Enhanced Differential Testing in Emerging Database Systems}

\author{Yuancheng Jiang, Jianing Wang, Chuqi Zhang, Roland Yap, Zhenkai Liang, Manuel Rigger}
\affiliation{%
  \institution{National University of Singapore}
  \country{Singapore}
  \city{Singapore}
}
\email{{yuancheng, jianing, chuqiz, ryap, liangzk, rigger}@comp.nus.edu.sg}

\begin{abstract}

In recent years, a plethora of database management systems have surfaced to meet the demands of various scenarios. 
Emerging database systems, such as time-series and streaming database systems, are tailored to specific use cases requiring enhanced functionality and performance. 
However, as they are typically less mature, there can be bugs that either cause incorrect results or errors impacting reliability. 
%
To tackle this, we propose enhanced differential testing to uncover various bugs in emerging SQL-like
database systems.
The challenge is how to deal with differences of these emerging databases.
Our insight is that many emerging database systems are conceptually extensions of relational database systems, making it possible to reveal logic bugs leveraging existing 
relational, known-reliable database systems.
%
However, due to inevitable syntax or semantics gaps, it remains challenging to scale differential testing to various emerging database systems.
We enhance differential testing for emerging database systems with three steps: (i) identifying shared clauses; (ii) extending shared clauses via mapping new features back to existing clauses of relational database systems; and (iii) generating differential inputs using extended shared clauses.
%
We implemented our approach in a tool called \name 
and applied it to four popular emerging database systems.
In total, we found \totalbug unknown bugs, of which \logicbug were logic bugs and \errorbug were internal errors. Overall, vendors fixed \fixbug bugs and confirmed \confbug.
%
Our results demonstrate the practicality and effectiveness of \name in detecting bugs in emerging database systems, which has the potential to improve the reliability of their applications.

\end{abstract}

\maketitle

\section{Introduction}

Recently, the landscape of database systems has witnessed a significant transformation with the emergence of diverse and specialized database systems (\textie emerging database systems).
These emerging database systems have been developed to address the evolving needs of modern applications and the complexities of handling large amounts of data.
Unlike traditional ones, emerging database systems are designed with a focus on specific use cases, offering optimized solutions for various industries and applications.
For instance, \textit{time-series} database systems (\texteg QuestDB~\cite{questdb}) 
are tailored for time-series data.
\textit{Streaming} database systems (\texteg Risingwave~\cite{risingwave}) differ in that they process streaming and in-memory queries with a low latency.

Emerging database systems, often developed under market pressure, are typically less mature and may have more bugs compared to established relational database systems. Database system testing aims to detect such bugs, which include internal errors and logic bugs. Internal errors are unexpected aborts, exceptions, or crashes in database systems, while logic bugs silently cause applications using these database systems to produce incorrect outcomes. While we detect both types of bugs, we place greater emphasis on detecting logic bugs. Unlike directly noticeable crash bugs, logic bugs often escape the attention of users and developers. Consequently, to effectively uncover these bugs, a proper test oracle is needed.

Despite the importance of finding bugs in emerging database systems, it has not garnered sufficient attention from both developers and researchers.
The only related work, Unicorn~\cite{wu2022unicorn}, aims to find internal errors in time-series database systems.
However, this approach, utilizing fuzzing techniques to trigger crashes, is specifically tailored for time-series database systems and lacks a suitable test oracle for finding logic bugs.
Moreover, its application is not versatile enough to be extended for bug finding across a variety of emerging database systems.
This limitation arises due to the distinct semantics and data representation inherent to each database system.
Such diversity calls for a more adaptable and comprehensive strategy to effectively identify and address more complex bugs in these varied environments.


In this work, our insight is that many emerging database systems can conceptually be seen as extensions of relational database systems, making it possible to reveal various bugs by reference to results from relational ones (\textie differential testing), which have been extensively tested~\cite{Rigger2020NoREC, Rigger2020PQS, rigger2020finding, zhong2020squirrel, liang2022detecting, ba2023testing, jiang2023dynsql} and are more robust.
To this end, we generate differential inputs based on clauses with the same syntax and semantics in both emerging and relational database systems, which we call \textsf{shared clauses}.
However, a practical challenge arises: differential testing is effective only for shared clauses, as differing syntax and semantics could produce false alarms. This limitation means that shared clauses may represent only a small fraction of dissimilar database systems' query languages, thereby restricting the effectiveness of differential testing in exposing bugs.

To tackle this limitation, we introduce a database system testing tool with enhanced differential testing, SQL-Cross-Differing (\name), to uncover bugs in SQL-like emerging database systems with the following steps: (i) clause identification, (ii) clause mapping, and (iii) differential inputs generation.
Specifically, \name first 
identifies supported clauses $C_1$ in the emerging (test) database system, and supported clauses $C_2$ in the relational (reference) database system. The common clauses, $C$=$C_1 \cap C_2$, form the \textsf{shared clauses}. Next, to realize our key insight, \name expands the shared clauses via mapping dedicated clauses in the test database system using existing clauses in the reference database systems (\textie obtain the \textsf{mapped clauses} $C_3$ and the extended set $C \cup C_3$). After we have a larger shared clause set, \name generates semantically equivalent, but syntactically different queries $Q_1$, $Q_2$ via randomly adding shared or mapped clauses for two target database systems. Such mapped pairs of queries are not guaranteed to be equal syntactically but should fetch the same result in different database instances. Therefore, if their results $R_1$ and $R_2$ are not equal, we detect a logic bug in the emerging database system given the trust in the maturity of relational database systems.

Listing~\ref{lst:motivate} demonstrates how our approach enhances differential testing.
The initial query $Q$ yields a different result in the test database system (QuestDB) from the reference database system (PostgreSQL). Nevertheless, this difference is expected after investigating their distinct treatments of \query{null}.\footnote{QuestDB regards \textsf{null} as a specific value and returns true because \textsf{null}(c0) is indeed in the list, while PostgreSQL returns \textsf{null} directly if the left expression is \textsf{null}}
Current differential testing methods either report such queries as being different, a false alarm, or ignore such features (\texteg avoid testing \query{null} values).
Our approach leverages clause mappings to address this gap. As shown in query $Q$(mapped), we map the \query{in} clause to a \query{case...when} clause. This allows us to first handle  \query{null} values and otherwise return the boolean value for non-null expressions. If the difference persists after mapping, we identify a bug-inducing case. 
In this case, QuestDB promptly acknowledged and resolved the issue.

\begin{figure}[t]
\setlength{\abovecaptionskip}{2pt}
\setlength{\belowcaptionskip}{0pt}
\setlength{\intextsep}{0pt}
\begin{lstlisting}[escapeinside={(*}{*)}, language=SQL, label=lst:motivate, caption=Motivating Logic Bug Found in QuestDB]
CREATE TABLE test (c0 INT); -- Initialize Schema --
INSERT INTO test VALUES (NULL); -- Initialize Data --
Q: SELECT (c0 IN (0, NULL)) FROM test; -- False Alarm --
-- QuestDB Result: [(True)] Postgres Result: [(Null)] --
Q(mapped): SELECT (CASE WHEN c0 IS NULL THEN NULL ELSE c0 IN (0) END) FROM test; -- True Fixed Bug --
-- QuestDB Result After Mapping: [(False)]!=[(Null)] --
\end{lstlisting}
\end{figure}

\sloppy
To assess the effectiveness of our technique, we used \name to test four popular emerging database systems: QuestDB~\cite{questdb}, TDEngine~\cite{tdengine}, RisingWave~\cite{risingwave}, and CrateDB~\cite{cratedb}. All selected emerging database systems have growing popularity (\texteg having at least thousands of stars in Github) and are actively maintained and updated. We found a total of \totalbug previously unknown bugs (\logicbug logic bugs and \errorbug internal errors) where \fixbug of them have been fixed and \confbug confirmed.
We evaluated \name on its improvements in bug finding for SQL-like emerging database systems:
(i) demonstrating its effectiveness in finding more unknown bugs; and (ii) greater coverage with more unique query plans.
We also compared \name against the state-of-the-art test oracle, Ternary Logic Partition (TLP)~\cite{rigger2020finding}, and its bug-finding tool SQLancer~\cite{Rigger2020PQS}.
The results show that \name is more effective (\texteg identified \logicbug logic bugs that SQLancer was unable to find) in detecting bugs and covers more unique query plans and greater code coverage when testing SQL-like emerging database systems.
Our efforts received encouraging acknowledgments from QuestDB developers in a post blog.\footnote{Acknowledged at \url{https://questdb.io/blog/fuzz-testing-questdb}} 


In summary, we make the following contributions:
\begin{itemize}
[topsep=0.2mm,parsep=0.2mm,partopsep=0pt,leftmargin=*]
\setlength{\itemsep}{0pt}
\setlength{\parsep}{0pt}
\setlength{\parskip}{0pt}
    \item We propose a key insight that emerging database systems are often extensions of relational ones, enabling extended differential testing through syntax mapping between database instances.
    \item We develop a practical tool called \name\footnote{\name will be available upon paper acceptance.}, which generates semantically equivalent queries expressed in different syntax by clause identification, clause mapping, and query generation.
    \item \name has found many bugs (logic and internal errors) in popular emerging database systems. Our bug reports were positively received and acknowledged by the developers.
\end{itemize}
\section{Emerging Database Systems}

\label{sec:background}

\begin{figure}[t]
\setlength{\abovecaptionskip}{2pt}
\setlength{\belowcaptionskip}{0pt}
\setlength{\intextsep}{0pt}
    \centering
    \includegraphics[scale=0.29]{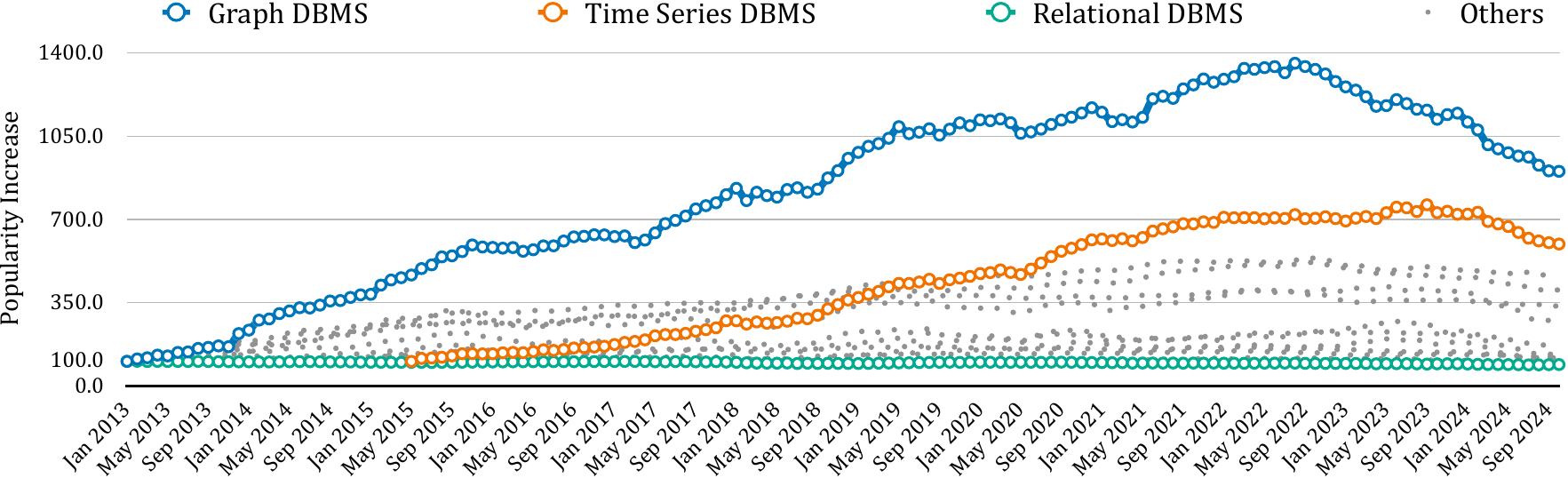}
    \caption{Popularity Trend of Database Systems Past Decade}
    \label{fig:trend1}
\end{figure}
\captionsetup{skip=0pt}

\vspace{-1mm}

In this section, we describe background knowledge about emerging database systems to (i) illustrate the importance of testing them, and (ii) give an intuition of our proposed testing methodology. 
Firstly, we examine the scope of these newly introduced database systems and study their popularity. 
Subsequently, we study the feature differences among these emerging database systems. We explain that their new features could be extended from existing expressions in relational database systems. 
Moreover, we explore their query languages and interfaces to identify that Structure Query Language (SQL) is the predominant query language.
%
In summary, we formulate and address the following research questions:

\begin{itemize}[noitemsep,topsep=0.2mm,parsep=0.2mm,partopsep=0pt,leftmargin=*]
\setlength{\itemsep}{1pt}
\setlength{\parsep}{0pt}
\setlength{\parskip}{0pt}
    \item[-] \textit{RQ1: What are the types of emerging database systems, and how is their popularity compared with relational database systems?}
    \item[-] \textit{RQ2: What are new features in emerging database systems?}
    \item[-] \textit{RQ3: What are query languages in emerging database systems?}
\end{itemize}
 
To systematically explore emerging database systems, we collect the latest statistics from reputable database survey platforms, including db-systems~\cite{dbengine} and the database of databases (dbdb)~\cite{dbdb}.
These platforms provide up-to-date information, including popularity scores, database types, and query interfaces (query languages) on newly introduced database systems. 
We also refer to their official documentation and open-source platforms (\texteg GitHub) for supplementary statistics and information.

\vspace{-1mm}
\paragraph{\textbf{RQ1: Types and popularity}.}
One notable trend in the evolution of database systems is the increasing popularity of emerging database systems, driven by their introduction of new features and specialized data models tailored to specific data types and use cases. However, these emerging database systems face the challenge of enhancing their accuracy and robustness through effective database system testing. In this paper, we categorize database systems into two groups: (a) conventional database systems and (b) emerging database systems. 


\textit{Conventional Database Systems}.
Conventional database systems typically refer to relational database systems, which have been well-established and extensively developed over several decades. They are the mainstream in database system adoption due to their long-standing reliability.
%
%
Well-known relational database systems include MySQL~\cite{mysql}, SQLite~\cite{sqlite}, and PostgreSQL~\cite{postgres}.
Numerous studies~\cite{Rigger2020NoREC, Rigger2020PQS, rigger2020finding, zhong2020squirrel, liang2022detecting, ba2023testing, jiang2023dynsql} have focused on testing relational database systems, demonstrating their effectiveness in identifying many previously unknown bugs.

\textit{Graph Database Systems}. Graph database systems such as Neo4J~\cite{Neo4j} and TinkerPop~\cite{TinkerPop} are the most popular emerging database systems in the past decade, which have experienced rapid growth, as they directly work on graph data.
Despite being newer compared to relational database systems, there has been considerable research on testing these systems~\cite{zheng2022finding, kamm2022testing, hua2023gdsmith, song2023testing, zhuang2023testing, jiang2024detecting, mang2024testing} to uncover unknown bugs.
We do not further discuss them, given these extensive existing efforts in creating effective testing approaches.

\textit{Emerging Database Systems}.
We refer to emerging database systems as those proposed with dedicated features within the last ten years and which have seen a notable increase in popularity, with more than a 10\% average rise in the period.\footnote{
This upward trend corresponds to the ``Slope of Enlightenment'' in the Gartner Hype Cycle~\cite{gartner}, reflecting an increasing acknowledgment of their practical usage.
}
Besides graph database systems, other types of emerging database systems are also growing fast. 
These database systems (\texteg time-series database systems, vector database systems, streaming database systems, wide-column database systems, and spatial database systems) introduce dedicated features that provide more convenient functionalities. In contrast to their widespread scale and popularity, there has been limited research into automated bug discovery \cite{wu2022unicorn, yang2023randomized}.

To better present their popularity differences, we collect statistics from the aforementioned survey sources~\cite{dbengine, dbdb} to summarize the popularity trends from 2013, with an initial popularity score of 100, of 
widely-used database systems during the past decade shown in Figure~\ref{fig:trend1}. The statistics reveal that graph database systems are leading database models in the past decade while relational database systems remain stable on the bottom. 
The statistics indicate that emerging database systems are experiencing rapid growth (\texteg time-series database systems are the first fastest-growing category except for graph ones, as shown in Figure~\ref{fig:trend1}). However, despite this growth, they surprisingly receive less attention in testing efforts compared to graph and relational database systems.

\vspace{-1mm}

\paragraph{\textbf{RQ2: New features}.} 
To effectively showcase the landscape of emerging database systems, we have opted to focus on two types of them. Namely, time-series database systems due to their growing popularity shown in Figure~\ref{fig:trend1}, and streaming ones for their increasing importance in handling timing-critical tasks.

\textit{Time-series database systems} have gained prominence in managing data points indexed by time, making them ideal for applications like Internet of Things (IoT) devices, financial trading, and monitoring systems. 
These database systems (\texteg QuestDB~\cite{questdb}, TDEngine~\cite{tdengine}) provide new features (\texteg date differing, date sampling) related to date and timestamp as well as storage optimization and retrieval of timestamp data, enabling efficient analysis and visualization of trends over time. Listing~\ref{lst:tsdb-demo} gives an example query in QuestDB, which utilizes special clauses, namely \query{sample by} and \query{fill}, to visualize the histogram of timestamp data. This query retrieves results from the table, identifying maximum and minimum values based on timestamps sampled at one-hour intervals. Additionally, it fills in the results based on the preceding one in case no results are available for a given duration. One more new feature is the clause \query{latest on}, which retrieves the most recent entry by timestamp for a given key or combination of keys. Additionally, time-series database systems include various timestamp functions or operations like \query{dateadd()}, \query{datediff()}, and \query{in}. Such new features provide a direct way to access or modify timestamped data.

\textit{Streaming database systems} like RisingWave~\cite{risingwave} are vital for applications demanding instant data updates and seamless synchronization across users and systems. These systems are used in applications needing streaming features, such as
messaging platforms, collaborative tools, and live analytics dashboards, which need
 low-latency access to data and concurrent user interactions.
To achieve this goal, streaming database systems introduce various new features. For instance, RisingWave provides a new window function called \query{tumble}. The time window refers to temporal intervals that users can utilize to segment events and execute data computations in the streaming process. The \query{tumble(table\_or\_source, start\_time, window\_size)} function in RisingWave generates a new table alternative for data selection. Additionally, it also supports window \query{join} that combines several time windows for convenient queries in streaming data. 

\begin{figure}[t]
\begin{lstlisting}[escapeinside={(*}{*)}, language=SQL, label=lst:tsdb-demo, caption=New Feature in QuestDB (Time-Series)]
SELECT ts,max(a),min(b) FROM T SAMPLE BY 1h FILL(PREV);
-- 2023-01-01T01:00:00.000000Z  max1  min1 --
-- 2023-01-01T02:00:00.000000Z  max1  min1 -- (filled)
-- 2023-01-01T03:00:00.000000Z  max3  min3 -- 
\end{lstlisting}
\end{figure}


\vspace{-0.1cm}

\paragraph{\textbf{RQ3: Query languages}.}
In light of varied data models proposed in various emerging database systems, the statistics from the ``database of databases''~\cite{dbdb} show a total of 18 distinct query interfaces (\texteg command line, SQL, custom API, Cypher, Datalog).
Among 450 individual database systems that have surfaced over the past decade, 158 instances (35.1\%) have opted for SQL-like queries as their interfaces. This places SQL-like query languages as the most common choice. Hence, this work focuses on testing emerging database systems with reference to SQL to provide wider applicability.

\begin{figure*}[t]
\setlength{\abovecaptionskip}{2pt}
\setlength{\belowcaptionskip}{0pt}
\setlength{\intextsep}{0pt}
    \centering
    \includegraphics[scale=0.63]{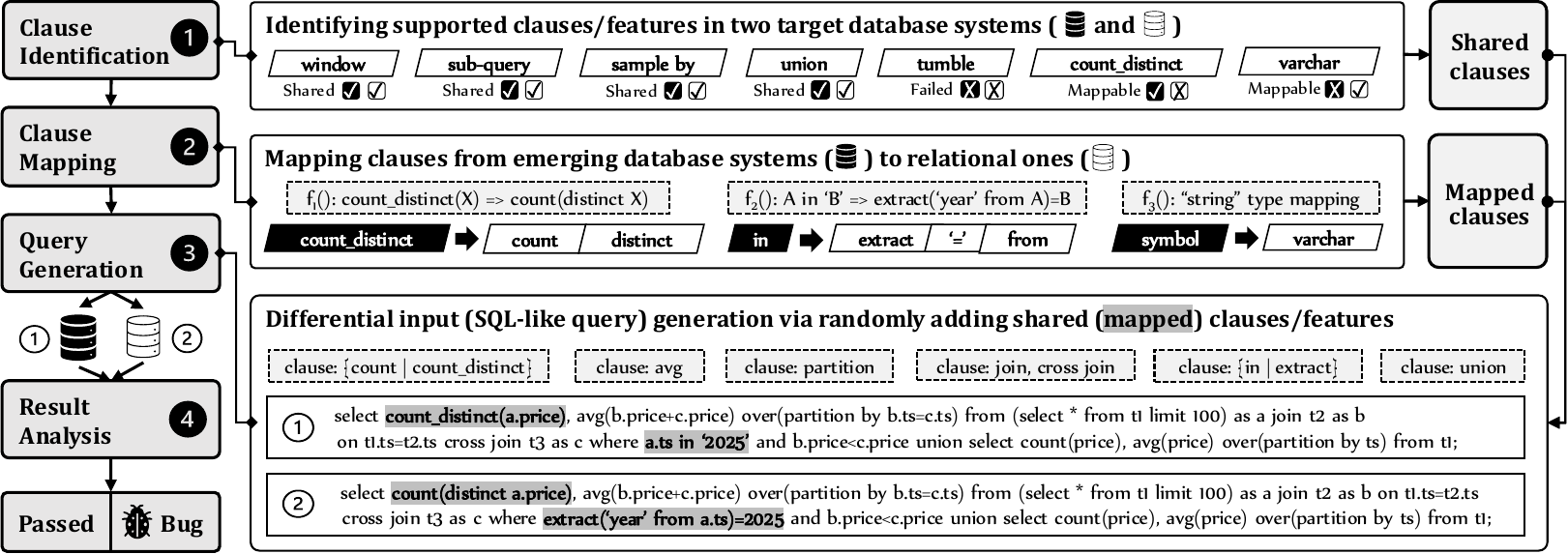}
    \caption{Approach Overview}
    \label{fig:overview}
\end{figure*}
\section{Approach}

\label{sec:app}

Our core insight is that many emerging database systems can conceptually be seen as an extension of relational database systems, making it possible to reveal logic bugs by checking whether the emerging database systems' results match the relational database systems' ones. We introduce enhanced differential testing that allows more extensive test cases to be evaluated on various emerging database systems. Our approach has three key steps: (i) identifying shared clauses; (ii) expanding the shared clauses by mapping dedicated clauses or features in emerging database systems back to relational ones; and (iii) generating (mapped) differential inputs through randomly selecting shared and mapped clauses. The results of the differential queries are then compared to determine logic bugs.
Given the widespread adoption and massive testing of relational database systems, we consider their results to be reliable and accurate. Therefore, differences in the query results suggest the presence of a logic bug in the emerging database systems. 

We believe that this simple approach relying on finding and utilizing commonalities in these related query languages has broad applicability across database systems supporting SQL-like query languages. 
The clause mapping is not expected to be general but necessarily specific to the kind of emerging database systems (or even a specific database system pair) under test. 
In this paper, we demonstrate the application and effectiveness of our approach on two prominent types of emerging database systems, specifically time-series and streaming database systems identified in Section \ref{sec:background}.


\paragraph{\textbf{Approach overview}.} Our methodology aims to generate semantically equivalent queries for emerging SQL-like database systems (the system under test) against established relational database systems (the reference result).
The basic premise of our test oracle is that semantically equivalent queries should yield identical results, even when they are syntactically different. Figure~\ref{fig:overview} uses the time-series database system QuestDB to illustrate our approach.

At step \circled{1}, \name initiates the testing process by determining supported clauses or features in the target and reference database systems. \name employs a series of tests on various clauses to categorize these features into three distinct groups: \textit{Shared}, \textit{Failed}, and \textit{Mappable}. When a clause successfully passes the tests in both database instances, it is labeled as \textit{Shared} and included in the set of shared clauses. Conversely, if a clause fails in both instances, it is labeled as \textit{Failed}, and the clause is discarded. In cases where a clause partially succeeds, it is designated as \textit{Mappable}, and we attempt to further map such clauses in the next step.


Step \circled{2} tries to expand the collection of clauses by using clause mappings to convert \textit{Mappable} clauses into appropriate representations in the reference database system. For example, we use $f_1$ to convert \query{count\_distinct(x)} into \query{count(distinct x)} to ensure compatibility with PostgreSQL's function syntax, and $f_2$ adjusts the usage of \query{in} in QuestDB to conform to valid SQL syntax for the \query{extract} function. 



At step \circled{3}, \name uses the expanded set of clauses (\textie shared clauses and mapped clauses) to generate valid queries for both database instances. Due to the introduced clause mappings in the preceding step, the generated query pair may be syntactically different, which we call \emph{mapped query generation}. For each shared or mapped clause, we employ random selections to incorporate such clauses. 
Figure~\ref{fig:overview} shows the process of generating mapped queries for testing QuestDB and PostgreSQL highlighting the mapped clauses. We then execute the queries on the test and reference database systems to retrieve the results.

\begin{table*}[t]
\caption{Illustrative List of Clause Mappings with Examples from Emerging to Relational Database Systems}
\centering
\begin{adjustbox}{width=0.99\textwidth}
\begin{tabular}{llll}
\toprule
\textbf{ID} & \textbf{Clause} & \textbf{QuestDB (01-08) / RisingWave (09-10)}  & \textbf{PostgreSQL Mapping} \\ \hline
01 & \tablequery{Sample By} & \tablequery{select count(*) from T sample by 1h} & \tablequery{select
a from (select count(*) as a, extract(hour from date) as b from T group by b)} \\
02 & \tablequery{Null} & \tablequery{select A in (1,2,3,null)} & \tablequery{select case when A is null then null else A in (1,2,3) end} \\ 
03 & \tablequery{DateAdd} & \tablequery{select dateadd(`h',1,ts)} & \tablequery{select cast((cast(ts as integer)+3600) as timestamp)} \\
04 & \tablequery{DateDiff} & \tablequery{select datediff(`y',now(),now())} & \tablequery{select abs(extract(year from now())-extract(year from now()))} \\
05 & \tablequery{Latest On} & \tablequery{select * from t latest on a partition by b} & \tablequery{.. (select .. join (select distinct max(a) over(partition by b) as a, b from t) ..} \\
06 & \tablequery{Distinct} & \tablequery{select count\_distinct(c0) from test} & \tablequery{select count(distinct c0) from test} \\
07 & \tablequery{Symbol} & \tablequery{create table test (c0 int, c1 symbol, c2 timestamp)} & \tablequery{create table test (c0 int, c1 varchar(128), c2 timestamp)} \\
08 & \tablequery{Between} & \tablequery{select count(*) from T where c0 between 2 and 0} & \tablequery{select count(*) from T where c0 symmetric between 2 and 0} \\
09 & \tablequery{Tumble} & \tablequery{select * from tumble(test, c0, interval `1 day')} & \tablequery{.. (select *, date\_trunc(`day',c0) as s .. interval `1 day') .. order by s, e) tumble} \\
10 & \tablequery{Hop} & \tablequery{.. hop(test, c2, interval `1 day', interval `2 days')} & \tablequery{.. (select *, date\_trunc(`day',c0) as s .. interval `2 day') .. order by s, e) hop} \\
\bottomrule
\end{tabular}
\end{adjustbox}
\label{table:mappings}
\end{table*}


Finally, \name performs result analysis at step  \circled{4}. Our test oracle assumes semantically equivalent queries should yield identical results even with syntax differences.
While either of the two systems might be affected by a bug, it is more likely that the emerging database system will be affected due to its lower maturity.
Upon reproducing and validating the discrepancies, we document and report the identified bugs to the respective developers. The key challenge of our approach is how to improve the extensibility of differential testing by extending clauses via clause mappings that transfer dedicated features in emerging database systems into valid expressions in relational database systems. 
The exact clause mappings are highly dependent on the emerging database system under test. We believe that developers of emerging database systems have a thorough understanding of how their systems' features differ from established relational database systems; thus, we expect that they can efficiently adopt our proposed approach. For illustration, we have selected two emerging database systems, time-series and streaming to give examples of mappings.






\paragraph{\textbf{Clause mapping for time-series database systems}.} From the statistics in Figure~\ref{fig:trend1}, time-series database systems are the second fastest-increasing database model in the past decade.
To test them, we first create a database schema with a timestamp column consistently set as the primary key for the table.
%
We demonstrate clause mappings for QuestDB, a fast-growing time-series database system that follows SQL syntax. Similar mappings exist for other time-series database systems. PostgreSQL is used as a reference for a reliable relational database system. In Table~\ref{table:mappings}, we present notable clause mappings along with corresponding query examples before and after the mappings. A detailed explanation follows.

(i) Mapping for clause \query{sample by}. From Section~\ref{sec:background}, the \query{sample by} is a new clause introduced in QuestDB to summarize large datasets into aggregates of homogeneous time periods as part of a select statement (\texteg to process histogram of timestamp data). Consider the query \query{select count(*) from sensors sample by 1h} in QuestDB, which retrieves record numbers of every hour. Relational database systems, however, do not support such keywords. 
We address this by translating it into the valid query \query{select sample\_by\_result from (select count(*) as sample\_by\_result, extract(hour from date) as hour from sensors group by hour)} in PostgreSQL via a combination of \query{group by} and sub-query to generate an equivalent query to \query{sample by}. 

(ii) Mapping for \query{null} values. The \query{null} value in QuestDB differs from most database systems, representing a specific value rather than the special nonexistent value in SQL~\cite{guagliardo2017correctness, guagliardo2017formal}. When a comparison involves \query{null},  QuestDB outputs boolean values while PostgreSQL outputs \query{null}. To bridge this gap, we make use of the shared clause \query{case...when} to first identify the \query{null} value in comparison operands, and if it exists, outputs \query{null} directly in QuestDB instead of boolean values. 
Consider the query \query{select A in (B)}. We map this query into \query{select case when A is null then null else A in (B) end} to align the result as \query{null} with PostgreSQL. Similar clause mappings exist in other comparison clauses (\texteg \query{between}). While this appears simple, it is effective in reducing the false alarms of differential testing.

(iii) Mapping for date operations including \query{dateadd}, \query{datediff}, and \query{in}. These functions provide more straightforward interfaces for directly operating on timestamp data.
For the \query{dateadd}, we first cast timestamp data to an integer and cast it back after calculations. For example, \query{dateadd(`h',1,ts)} is translated into \query{cast((cast(ts as integer)+3600) as timestamp)}. 
For \query{datediff}, the query \query{datediff(`y',now(),now())} in QuestDB returns the result \query{0}, indicating that the two dates differ by zero years. 
We adopt clause mappings in our approach and translate it into \query{abs(extract(year from now())-extract(year from now()))} as a valid and semantically equivalent query in PostgreSQL. 
The \query{in} operator checks whether the given data is within the range. With timestamp data, it can be used as \query{ts in `2023'} to check if the timestamp is in the year 2023. We adapt it to relational database systems via the clause \query{between}. For example,  \query{ts in `2023-01;-3d'} can be translated to \query{ts between `2023-01-01 00:00:00' and `2023-01-28 23:59:59'}. Figure~\ref{fig:overview} also gives a complete example showing the clause mapping of date operations from the target database system to reference one.

(iv) Mapping for clause \query{latest on}. The \query{latest on} clause in QuestDB retrieves the most recent data row by timestamp for given keys. Consider the query \query{select * from t latest on c0 partition by c1} where \query{c0} represents the designated timestamp primary key and \query{c1} is an integer data column. 
PostgreSQL does not support this feature. 
We map it into \query{select distinct * from (select t1.c0,t1.c1,t1.c2,t1.c3 from t as t1 join (select distinct max(c0) over(partition by c1) as c0, c1 from t) as t2 on t1.c0=t2.c0 and t1.c1=t2.c1) latest\_on} by replacing the \query{latest on} with table joins and a
window function. 

(v) In addition to dedicated features in emerging database systems, clause mappings are essential when common clauses have different semantics or syntax in two database instances. For example, the clauses \query{between} and \query{symmetric between} are both used to select values within a given range (\textie between the lower and upper bounds, inclusively). Some database systems, like QuestDB, implicitly use \query{symmetric between} to accommodate invalid ranges, while PostgreSQL explicitly requires the \query{symmetric} keyword.
Another crucial aspect of clause mapping is type aliasing, which is necessary when working with various database systems that use different type names. To ensure consistency and prevent syntax errors, we employ data type alias mappings including \query{symbol} to \query{varchar} or \query{text}, \query{int} to \query{integer}, \query{long} to \query{bigint}, and \query{timestamp} to \query{datetime}.
In some cases, clause mappings help connect similar clause usages. For example, QuestDB has special functions for distinct aggregation, such as \query{count\_distinct(x)}, while PostgreSQL uses the \query{distinct} keyword inside the count function as \query{count(distinct x)}. Although these clause mappings may seem straightforward, they are vital for accurate differential testing across dissimilar database systems.

\vspace{-1mm}

\paragraph{\textbf{Clause mapping for streaming database systems}.} Streaming database systems are designed for handling streaming data with optimizations and advanced features. To deal with streaming data, these systems process information as soon as it arrives, rather than waiting until it has been stored. We show parts of clause mappings from Table~\ref{table:mappings} in the streaming database system RisingWave~\cite{risingwave} as follows: (i) Mapping for time window function \query{tumble}. 
In RisingWave, the \query{tumble} function generates a new table alternative for data selection. 
Nevertheless, relational database systems do not support such features. 
We translate \query{tumble} into a sub-query in relational database systems with additional columns window\_start and window\_end. Take the \query{select * from tumble(t, c0, interval `1 day')} as an example. We translate it into a valid query \query{select * from (select *, date\_trunc(`day', c0) as s, date\_trunc(`day', c0 + interval `1 day') as e from t order by s,e) tumble\_table} in PostgreSQL. (ii) Similar to the \query{tumble} function, the \query{hop} function creates a new table with additional consideration of time hop windows. We utilize similar translations through sub-queries with adjusted intervals. 
(iii) Mapping for \query{window join}: The \query{window join} refers to joining a time window with a table or another time window of the same type. As we map \query{tumble} or \query{hop} into sub-queries in (i) and (ii), we intuitively join these two sub-queries to achieve an equivalent query in PostgreSQL. (iv) Other similar clause mappings (\texteg type mapping) in QuestDB are also required in RisingWave. For brevity, we omit clause mappings in RisingWave that correspond to those in QuestDB.

\begin{figure}[t]
\setlength{\abovecaptionskip}{2pt}
\setlength{\belowcaptionskip}{0pt}
\setlength{\intextsep}{0pt}
    \centering
    \includegraphics[scale=0.5]{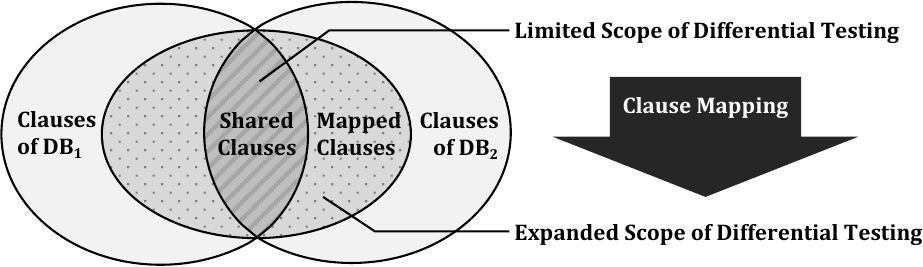}
    \caption{How Our Approach Improves Differential Testing}
    \label{fig:dt_improve}
\end{figure}

\paragraph{\textbf{Discussion of clause mapping}.}
In this paper, we propose the use of clause mappings to simplify the problem of testing different SQL-like databases with differences in syntax and semantics.
One significant advantage is that we employ clause mappings to 
enhance testing coverage of database queries by translating clauses between database systems, potentially improving overall testing effectiveness.
Figure~\ref{fig:dt_improve} illustrates that shared clauses found may be only a small fraction while expanding with mapped clauses extends the range of generated queries.
We believe this approach is highly scalable and can be applied to various other database engines.

Deriving clause mappings requires some manual effort to explore the documentation and conduct mapping experiments on database systems.
We highlight that manual effort of some form is common among database system testing works. For example, as a metamorphic testing approach, SQLancer~\cite{Rigger2020PQS} requires considerable manual efforts, around 3K lines of Java code for each supported DBMS. Manual effort is needed simply because even with SQL databases, SQL dialects differ in both syntax and semantics which is not dissimilar to the need for manual effort in \name.
Another example is Thanos~\cite{fu2025thanos}, which uses differential testing. It requires manual effort to collect supported features from various storage engines.

In \name, our manual effort is a reasonable one-time cost and an easy task for developers of emerging database systems, as they have likely developed their systems' new features after studying the functionalities of existing established database systems. 
Based on our experience, 10 to 20 clause mappings are already sufficient to identify various bugs, and implementing a clause mapping takes minutes to hours (\textie a few days to adapt to one emerging DBMS). 
Furthermore, it may be possible to automatically obtain clause mappings in the future using large language models.
Our contribution lies in demonstrating the feasibility of applying differential testing to diverse database systems, thereby enhancing the extensibility of differential testing.

Clause mappings do not need to be complete to enhance differential testing, as shown in Figure~\ref{fig:dt_improve}, differential testing already handles shared clauses and we enlarge the working scope of differential testing. Partial clause mappings (\texteg 10--20 clause mappings) significantly expand test generation. We highlight that none of related works~\cite{Rigger2020PQS, rigger2020finding} considers the completeness of database system testing. Rather, an approach is considered effective if it finds bugs.


\captionsetup{skip=0pt}

\section{Implementation}


\label{sec:implementation}

We present the implementation of \name as shown in Figure~\ref{fig:overview}:
(i) \textit{clause identification}, (ii) \textit{database creation} (iii) \textit{query generation}, and (iv) \textit{result analysis}. Note that while some aspects are not the key contribution of our work, they still significantly contribute to effective and extensible differential testing.


\paragraph{\textbf{Clause identification}.} Clause identification aims to automatically uncover supported SQL clauses in the given target database system. We continuously collect and complement clause candidates (to be tested via a simple query) over three months by manually analyzing official documentation from various database systems with the assistance of large language models~\cite{ChatGPT}. 
In addition to various clauses, our approach also tests SQL features (\texteg sub-query) to ensure the differential inputs reach more complex and deep logic.

\textbf{\textit{Database creation}.} Before each round of testing, we initialize three tables. These tables can have up to 8 columns, each assigned a data type such as integer, string, or timestamp, which we populated with 50 rows to 500 rows. The data for these tables is generated randomly, with special values (\texteg \query{null}) taken into consideration. 

\textbf{\textit{Query generation}.} 
We generate SQL statements based on the three initialized tables by randomly adding shared or mapped clauses derived from our main approach. Our query generation is modular and scalable, allowing users to enable or disable various clauses or features as needed. We use several strategies to improve the complexity of generated queries: (i) replacing tables or expressions with sub-queries. Sub-queries are an important feature widely used in SQL that increases query complexity; (ii) using query concatenations. Query concatenations (\textie \query{union}, \query{except}, and \query{intersect}) can combine multiple queries into a single, more complex query; and (iii) add advanced features. Advanced features like window functions and branching are also included in query generation.



\textbf{\textit{Result analysis}.} The test oracle of \name is that we expect that query pairs constructed to generate the same results when executed on emerging and relational database systems indeed compute the same result. Otherwise, it suggests a logic bug.
Besides uncovering logic bugs, \name also detects internal errors. The main step is to distinguish whether one given exception is expected. We leverage heuristics to match the unexpected keywords (\texteg \query{NullPointerException}, \query{OutOfBoundException}) and exclude valid exceptions (\texteg feature not supported).
\section{Evaluation}



In this section, we answer the following questions to assess various important aspects of \name:

\begin{itemize}[noitemsep,topsep=0.2mm,parsep=0.2mm,partopsep=0pt,leftmargin=*]
\setlength{\itemsep}{1pt}
\setlength{\parsep}{0pt}
\setlength{\parskip}{0pt}
    \item \textbf{Q1 Discovery of Unknown Bugs}. 
    How effective is \name in discovering previously unknown bugs in popular emerging database systems?
    \item \textbf{Q2 Improvement on Differential Testing}. 
    To what extent does \name enhance the effectiveness of differential testing?
    \item \textbf{Q3 Comparison with Existing Techniques}. 
     What is the improvement in testing emerging database systems compared with the state-of-the-art database testing approaches?
    
\end{itemize}

\paragraph{\textbf{Selected database systems}.} To assess the efficacy of our approach, we have selected a subset of emerging SQL-like database systems, namely QuestDB~\cite{questdb} (14.5K stars), TDEngine~\cite{tdengine} (23.3K stars), RisingWave~\cite{risingwave} (6.9K stars), and CrateDB\footnote{CrateDB has multiple database models including time-series, geospatial, and vector.}~\cite{cratedb} (4.1K stars). 
Emerging database systems have a diverse range of types and models as shown in Section~\ref{sec:background}, making it challenging to cover all of them in this work. The selection is based on their increasing popularity and similarity to standard SQL syntax. We deliberately avoid including emerging database systems that are implemented as close extensions of well-known relational ones (\texteg Timescale~\cite{timescale}, closely mirrors PostgreSQL syntax, thus differential testing is trivially applicable). For relational database systems, we leverage PostgreSQL~\cite{postgres} as the reference ground truth for testing. Throughout the evaluation process, we spent most of our efforts on testing QuestDB, which differs significantly from PostgreSQL, having both newly-introduced features and semantic differences. 
We also applied our approach to other emerging database systems to demonstrate wide applicability, uncovering 6 bugs in TDEngine, 4 bugs in RisingWave, and 3 bugs in CrateDB, but we have spent less testing effort for those database systems.

\paragraph{\textbf{Experimental setup}.} We performed all experiments on a personal computer with Intel(R) Core(TM) i7-14700 CPU and 32GB RAM. The OS is Ubuntu 20.04.2 LTS. \name can efficiently detect all confirmed or fixed bugs within six hours of running on personal computing resources.

\begin{table}[t]
\centering
\caption{Unknown Bugs \name Found}
\begin{adjustbox}{width=0.47\textwidth}
\begin{tabular}{c|ccc|ccc|c}
\toprule 
& \multicolumn{3}{c|}{\textbf{Internal Errors}} & \multicolumn{3}{c|}{\textbf{Logic Bugs}} & \textbf{In Total} \\ \hline
\textbf{Database System} & \textbf{Unknown}    & \textbf{Confirmed}    & \textbf{\xspace\xspace\xspace Fixed \xspace\xspace\xspace}    & \textbf{Unknown}   & \textbf{Confirmed}  & \textbf{\xspace\xspace\xspace Fixed \xspace\xspace\xspace}  & -        \\ \hline
\textbf{QuestDB}         & 0          & 0            & 32         & 1         & 2          & 9      & \textbf{44}       \\
\textbf{TDEngine}        & 0          & 1            & 2        & 1         & 1          & 1      & \textbf{6}        \\
\textbf{RisingWave}      & 0          & 0            & 3        & 0         & 1          & 0      & \textbf{4}        \\
\textbf{CrateDB}         & 0          & 0            & 2        & 0         & 0          & 1      & \textbf{3}        \\ \hline
\textbf{In Total}        & \multicolumn{3}{c|}{\textbf{40}}      & \multicolumn{3}{c|}{\textbf{17}}         & \textbf{57} \\ 
\bottomrule
\end{tabular}
\end{adjustbox}
\label{table:bugs}
\end{table}

\subsection{Discovering Unknown Bugs}
\label{sec:bugs}

To detect new logic bugs in emerging database systems, we intermittently tested the latest versions of the target emerging database systems, using PostgreSQL-14.0 to derive the ground-truth results, over a period of three months, which is a typical methodology for evaluating the effectiveness of automatic testing tools~\cite{Rigger2020PQS, kamm2022testing}. In most cases, \name took a few hours until a bug was found under personal computer resources. 
We reported bugs after reducing bug-inducing queries and checking whether the issue had already been reported on issue trackers to avoid duplicate bug reports. 
Bug-inducing test cases generated automatically are typically complex, and we reduced them to a smaller bug-inducing version by delta debugging~\cite{zeller2002simplifying}. Next, we illustrate bugs with reduced queries, omitting unnecessary query mappings.


\paragraph{\textbf{Results}.} Table~\ref{table:bugs} summarizes the number of previously unknown bugs identified using our approach. We classified the identified bugs into two distinct categories: \textit{(i) Internal Errors} refer to bugs where a query caused unexpected aborts, exceptions, or crashes in the target database system; \textit{(ii) Logic Bugs} refer to bugs found through discrepancies flagged by the differential testing. Additionally, we categorized all bugs into three statuses: \textit{(i) Unknown} bugs are those that have been identified and submitted but are awaiting further investigation by developers to determine the root cause; \textit{(ii) Confirmed} bugs are those that have been acknowledged by developers but have not yet been fixed; \textit{(iii) Fixed} bugs are those that have been confirmed and subsequently fixed by the developers.

\begin{figure}[b]
\begin{lstlisting}[escapeinside={(*}{*)}, language=SQL, label=lst:string_bug, caption=Incorrect String Comparison in QuestDB]
CREATE TABLE test (c_0 SYMBOL);
INSERT INTO test VALUES ('A');
SELECT count_distinct(c_0) FROM test WHERE c_0>'Z';
-- Emerging Database System Result: [(1)] --
CREATE TABLE test (c_0 VARCHAR(16));
INSERT INTO test VALUES ('A');
SELECT count(DISTINCT c_0) FROM test WHERE c_0>'Z';
-- Relational Database System Result: [(0)] --
\end{lstlisting}
\end{figure}

In total, we identified \totalbug unknown bugs (\logicbug logic bugs and \errorbug internal errors), of which \confbug were confirmed and \fixbug were fixed. For illustration, in the subsequent paragraphs, we present parts of noteworthy bugs that \name identified, describing them based on their root cause through our analysis and developers' feedback. 

\begin{figure}[b]
\begin{lstlisting}[escapeinside={(*}{*)}, language=SQL, label=lst:bug1, caption=Incorrect Concatenated Query in QuestDB/CrateDB]
(SELECT 1 UNION ALL SELECT 1) EXCEPT (SELECT 0);
(SELECT 1 UNION ALL SELECT 1) INTERSECT (SELECT 1);
-- QuestDB: [(1),(1)]   PostgreSQL: [(1)] --
CREATE TABLE test (c_0 TIMESTAMP, c_1 INT, c_2 FLOAT);
INSERT INTO test VALUES (946702800000, 8, 3.0);
INSERT INTO test VALUES (946688400000, 9, 7.0);
INSERT INTO test VALUES (946695600000, 6, 8.0);
(SELECT .. FROM test as t1 CROSS JOIN ...) UNION (...);
-- CrateDB: [(3,1,1), ..]   PostgreSQL: [(3,2,1), ..] --
\end{lstlisting}
\end{figure}


 


\paragraph{\textbf{Logic bug---incorrect string comparison}.} 
A common feature of database systems is string comparisons, which are commonly supported by either using specific functions like \query{strcmp} in MySQL or directly via operators (\texteg >,<,=). In QuestDB, we observed one discrepancy\footnote{\url{https://github.com/questdb/questdb/issues/3828}} that occurred when counting distinct results with string comparison in the predicate, as shown in Listing~\ref{lst:string_bug} from PostgreSQL.
Identifying this bug requires syntax-different differential inputs by mapping clauses to string-type keywords (\textie from \query{symbol} to \query{varchar}) and the clause \query{count}. This example illustrates the simplicity of clause mapping and the enhanced effectiveness it brings to differential testing.

\paragraph{\textbf{Logic bug---incorrect query concatenation}.} The \query{intersect} and \query{except} clauses are commonly used to combine results from two sub-queries.
We found a bug\footnote{\url{
https://github.com/questdb/questdb/issues/3580
}} shown in Listing~\ref{lst:bug1} which concerns the logic of \query{intersect} and \query{except} clauses in QuestDB.
The identification of this bug results from applying query concatenations which combine three queries into a result unit. Another similar bug was found when executing complex SQL structures with multiple joins and the same column names in CrateDB , as shown in Listing~\ref{lst:bug1}. We observe the difference in the result set from PostgreSQL after the necessary type mapping (\texteg to cast \query{timestamp} into \query{bigint} for comparison) when executing \query{union} queries with duplicated output columns. These bugs were resolved after we submitted the cases causing the bug to the developers. 

\paragraph{\textbf{Logic bug---incorrect nested joins and window function}.} Complex query structures (\texteg with nested joins or window functions) tend to trigger bugs in emerging database systems. One bug \name found is related to nested joins. A nested join in SQL involves the use of multiple \query{join} operations within a single query, which allows for the retrieval of data that spans multiple tables.
Listing~\ref{lst:bug-join} shows Q1, 
a bug-inducing query computing an incorrect result due to an unknown bug
\footnote{\url{https://github.com/questdb/questdb/issues/4010}} 
The issue is in the transitive filter pass when handling nested table joins.
Another bug that \name found is related to the window function, which calculates their results across a set of rows related to the current row. Unlike regular aggregate functions, window functions do not collapse the result set into a single value for each group. In QuestDB, we observe one bug-inducing case,\footnote{\url{https://github.com/questdb/questdb/issues/3936}} shown as Q2 in Listing~\ref{lst:bug-join}, which outputs differently compared to PostgreSQL when executing window function queries with \query{order by} clause and table joins. 

\paragraph{\textbf{Internal errors---incorrect syntax errors}.} As one category of internal errors, \name also detects various incorrect syntax errors through expanded differential testing. We identified these bugs by observing syntax errors in QuestDB while the same queries returned correctly in PostgreSQL. As shown in Listing~\ref{lst:abort_bugs}, the query Q1 or Q2 reports \query{invalid column c0/c1} in QuestDB, whereas the same query executes normally in PostgreSQL. These discrepancies found via differential testing help identify such unexpected syntax errors more efficiently. 

\paragraph{\textbf{Internal errors---core dumped/out of bound}.} Another notable category of bugs we encountered involves internal errors where the target database systems crashed. As shown in Listing~\ref{lst:error_bugs}, \name found a \query{segmentation fault} in TDEngine and an \query{out-of-bound} error in CrateDB. To date, \errorbug internal errors have all been confirmed or fixed by the developers.

\paragraph{\textbf{Discussion---false alarms}.} We do not observe any false alarms when running SQLxDiff. However, we encountered 3 false alarms while developing SQLxDiff. Note that any potential false alarms, as determined by us or the DBMSs' developers, can be easily addressed by modifying or removing the clause mapping. This is a common methodology and has been used in many other differential testing approaches~\cite{mckeeman1998differential, slutz1998massive, brumley2007towards, chapman2011automated, chen2016coverage, jung:apollo, zheng2022finding}.

\begin{figure}[t]
\begin{lstlisting}[escapeinside={(*}{*)}, language=SQL, label=lst:bug-join, caption=Logic Bugs with Nested Joins or Window Functions]
CREATE TABLE t(c0 TIMESTAMP, c1 INT, c2 INT);
INSERT INTO t VALUES('2025-01-01 10:00:00+00', 1, 1);
INSERT INTO t VALUES('2025-01-01 10:00:00+00', 1, 1);
Q1: SELECT count(1) FROM t as T1 CROSS JOIN t as T2 WHERE T2.c0='2000-01-01 00:00:00+00' INTERSECT SELECT count(1) FROM t as T1 JOIN t as T2 on T1.c0=T2.c0 JOIN t as T3 ON T2.c0=T3.c0;
-- QuestDB Result: [(0)] PostgreSQL Result: [] --
Q2: SELECT avg(T1.c_1) OVER(PARTITION BY 1=1 ORDER BY T1.c_0) FROM test AS T1 CROSS JOIN test AS T2;
-- QuestDB: [(1),(1.333333)] PostgreSQL: [(1),(1.5)] --
\end{lstlisting}
\end{figure}

\begin{figure}[t]
\begin{lstlisting}[escapeinside={(*}{*)}, language=SQL, label=lst:abort_bugs, caption=Unexpected Syntax Errors in QuestDB]
CREATE TABLE test (c0 short, c1 timestamp); -- QuestDB
CREATE TABLE test (c0 float, c1 timestamp); -- Postgres
Q1: SELECT avg(c0) FROM test UNION SELECT DISTINCT avg(c0) FROM test;
-- QuestDB: "Invalid Column c0" PostgreSQL: [()] --
Q2: SELECT count(1) FROM test AS T1 JOIN test AS T2 ON T1.c1<T2.c1 JOIN test AS T3 ON T2.c1=T3.c1;
-- QuestDB: "Invalid Column c1" PostgreSQL: [()] --
\end{lstlisting}
\end{figure}

\subsection{Improvement on Differential Testing}

One key question is how much we can improve upon traditional differential testing. We evaluate the enhanced effectiveness of our differential testing approach through three main aspects: (i) discovering more bugs, (ii) covering more unique query plans, and (iii) achieving higher query execution success rates.

\paragraph{\textbf{Enhanced effectiveness in detecting logic bugs}.} \name identifies significantly more logic bugs, benefiting from the clause mapping technique, which broadens the scope of differential testing by bridging non-shared clauses in the two target database systems. To evaluate the impact of clause mappings, we first conducted a manual analysis of all \logicbug logic bugs identified by \name to determine the necessity of these mappings for triggering the bugs. Following this, we compared the clause success rates with and without the use of clause mappings.

\begin{figure}[t]
\begin{lstlisting}[escapeinside={(*}{*)}, language=SQL, label=lst:error_bugs, caption=Internal Errors in TDEngine/CrateDB]
CREATE TABLE test(c0 TIMESTAMP);
insert into test values ('2025-01-01 00:00:00.000');
Q1: select count(1) from test as a join test as b on a.c0=b.c0 and a.c0 is null;
-- TDEngine: FATAL crash signal...(core dumped) --
Q2: SELECT DISTINCT count ... sys.summits as T2 JOIN sys.summits as T3 ON T2.mountain>T3.mountain;
-- CrateDB: Index 2 out of bounds --
\end{lstlisting}
\end{figure}

Among \logicbug logic bugs identified by \name, 7 of them are detected with the assistance of clause mappings. The necessary clause mappings for reproducing logic bugs include (i) type mappings, as type keywords may vary across different database instances (\texteg the string type may be represented as \query{symbol} or \query{varchar}). Without type mappings, the bug cannot be reproduced as bug-inducing cases (\texteg Listing~\ref{lst:string_bug}) would fail at the \query{create table} statement; (ii) function mappings, where functions like \query{count\_distinct()} in emerging database systems require rewriting as \query{count(distinct)} for correct functionality; and (iii) feature mappings, which align new features with valid keywords in relational database systems as introduced in Section~\ref{sec:app}. 
However, not every logic bug necessitates reproduction via clause mappings. For instance, the inaccurate intersect/except results in Listing~\ref{lst:bug1} are deemed valid in both emerging and relational database systems even without the aid of mappings. 
Our methodology expands differential testing by bridging syntax gaps and mapping new features, allowing many SQL queries to be valid in both systems. Our findings highlight the benefits of clause mappings in detecting bugs across different database models. 

\paragraph{\textbf{Covering more unique query plans}.} A query plan outlines the ordered steps (\texteg table scans, table joins) that database systems will take to access data. Created by the query optimizer, it aims to find the most efficient way to execute the query based on factors such as the database schema, data distribution, and available indexes. A higher number of executed query plans increases the probability that the testing tool can uncover unknown bugs, as demonstrated in existing works~\cite{ba2023testing, ba2024keep}.

\begin{figure}[b]
    \centering
    \includegraphics[scale=0.45]{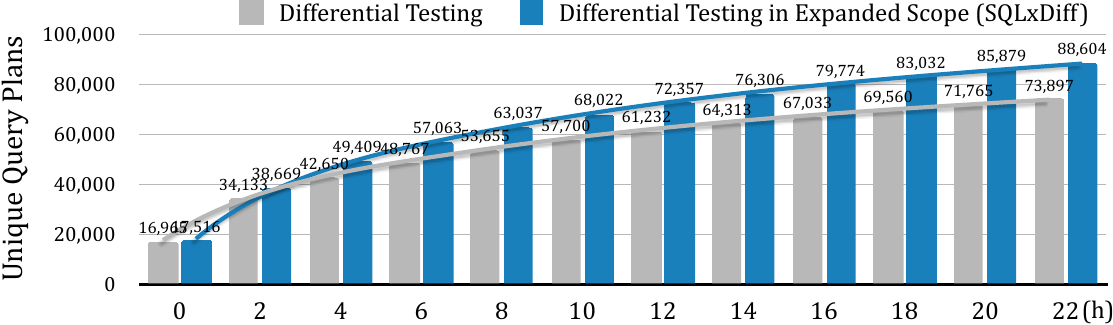}
    \caption{\name's Improvement on Unique Query Plans}
    \label{fig:dt_queryplan}
\end{figure}

We conducted a 24-hour experiment using \name to compare the unique query plans covered by classic differential testing (\textie without clause mapping) and an expanded scope of differential testing (\textie \name, with clause mapping). As shown in Figure ~\ref{fig:dt_queryplan}, \name covers 14,704 additional unique query plans, representing a 20\% improvement after 24 hours of testing. These results suggest that the expanded scope of differential testing improves the ability to execute a greater variety of query plans, thereby uncovering more unknown issues compared to traditional differential testing.

\paragraph{\textbf{Higher query execution success rate}.} To better illustrate the impact of clause mappings on the testing effectiveness of differential execution in both emerging database systems and relational database systems, we compare clause execution success rates with and without clause mappings. The success rates of queries exhibit significant variations under different configurations. We noted instances where the success rate dropped to 0\% when type mappings were not enabled, resulting in failures during table initialization. Conversely, the success rate remained stable when new features were infrequently incorporated into query generations. In summary, we observed a decrease in query success rates ranging from around 23.9\% (\texteg only queries with new features output syntax errors) to 100\% (\texteg all queries failed without necessary type mappings in \query{create table} statements) on QuestDB, RisingWave, CrateDB, and PostgreSQL. This underscores how clause mappings contribute to the efficacy of differential testing in dissimilar database systems.


\subsection{Comparison with Existing Techniques}

We now evaluate the effectiveness of \name compared with existing techniques.
To the best of our knowledge, there is only one (non-public) existing work on testing emerging database systems, Unicorn~\cite{wu2022unicorn}, which focuses on fuzzing time-series database systems. We contacted the authors to ask whether they could share their artifacts with us for comparison but have not received a response. Unicorn proposed hybrid input generation to create valid SQL queries for time-series database systems. 
However, as Unicorn uses a crash oracle, it is unable to 
detect more complex bugs like unexpected syntax errors and logic bugs. 

To further assess \name, we ported the state-of-the-art test oracle on relational database systems, Ternary Logic Partition (TLP)~\cite{rigger2020finding}
\footnote{Reasons for using TLP include: TLP is the state-of-the-art test oracle for finding logic bugs in SQL databases. TLP can be seen as a generalization of Non-Optimizing Reference Engine Construction (NoREC). Query Plan Guidance (QPG) is a test-case generation approach, and not a test oracle. Pivoted Query Synthesis (PQS) is no longer maintained in SQLancer and thus not used for comparison. Differential Query Execution (DQE) aims at testing Data Manipulation Language (DML) statements, which are out of the scope of this paper.}
implemented in SQLancer~\cite{Rigger2020PQS}, on our selected emerging database systems. Different from our approach, TLP leverages metamorphic mutations that partition one query into three sub-queries, each selecting rows based on whether a predicate generates true, false, or null. We compare the effectiveness by checking how many unknown logic bugs \name found can be detected by using the TLP test oracle. Then, we evaluate \name against SQLancer by comparing unique query plans and code coverage, demonstrating our improvements in testing emerging database systems.

\begin{figure}[t]
\begin{lstlisting}[escapeinside={(*}{*)}, language=SQL, label=lst:TLP1, caption=Consistent Sum Results---Ineffectiveness of TLP]
-- Buggy Query QuestDB Result [] vs. PostgreSQL [(1)]--
(SELECT DISTINCT avg(c0) over(partition by 1) FROM t);
-- TLP-True Query QuestDB Result [] --
(SELECT .. over(partition by 1) FROM t WHERE c0>0);
-- TLP-False Query QuestDB Result [] --
(SELECT .. over(partition by 1) FROM t WHERE not c0>0);
-- TLP-Null Query QuestDB Result [] --
(SELECT .. over(...) FROM t WHERE c0>0 is NULL);
\end{lstlisting}
\end{figure}

\begin{figure}[b]
    \centering
    \includegraphics[scale=0.45]{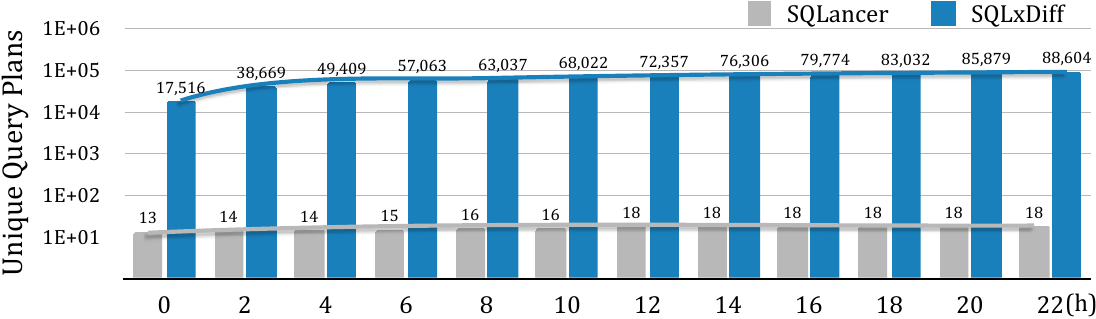}
    \caption{Comparison of Unique Query Plans}
    \label{fig:compareQP}
\end{figure}

\paragraph{\textbf{Effectiveness}.} We used the same methodology as prior works~\cite{rigger2020finding, Rigger2020NoREC}, that is, to conduct a manual and best-effort analysis to identify bugs found by \name that TLP overlooked. To adapt TLP in emerging database systems, we implement TLP to partition the original queries as a manual best-effort process leveraging existing predicates or introduce relevant predicates that refer to columns in the original queries in cases where no predicates exist. 



For \logicbug bug-inducing test cases to which test oracles could be applied, TLP does not surprisingly reveal any of these bugs, showing the unique strength of differential testing across database systems. The primary reason why TLP fails to detect them is that the presence of predicates does not influence the incorrect outcome. The root causes of such bugs are traced back to other clauses, such as \query{over(partition by)}, \query{join}, \query{union}, or unrelated expressions like \query{null} and features. In cases where these bugs originate, the conceptual addition of predicate partitions does not impact the results. A specific example is illustrated in Listing~\ref{lst:TLP1}. The bug initially occurs within a sub-query (enclosed in brackets) that fails differential testing in comparison with PostgreSQL. Upon query reduction, the bug-inducing scenario reveals that introducing brackets to a query with a window function causes the bug. Although we add predicates related to the sole column reference in the query, it does not influence the result, indicating that TLP fails to find this bug.

In addition to logic bugs, \abortbug of \errorbug internal errors \name found output inconsistent aborts, where the differential input correctly returns in the reference database system. Detecting such bugs generally requires more than a simple test oracle of metamorphic testing; it usually necessitates additional handlers for errors, exceptions, or crashes (\texteg Unicorn~\cite{wu2022unicorn}). In contrast, our enhanced differential testing 
leverages the correct output of the reference database system, enabling efficient identification of these bugs.

\paragraph{\textbf{Unique query plans}.} We conducted comparison experiments against SQLancer, evaluating the number of unique query plans observed over 24 hours to simulate the bug-finding capabilities of various testing approaches. Although SQLancer supports QuestDB, this support is only in its initial stages. We also found open bugs\footnote{\url{https://github.com/sqlancer/sqlancer/issues/712}} in its issue tracker, which we encountered and fixed, aiming for a fairer comparison. 
We ensured that each query plan was unique by inspecting the sequence and order of steps involved.

As shown in Figure~\ref{fig:compareQP}, \name demonstrated significant improvements compared to SQLancer. The statistics indicate that even state-of-the-art testing tools face challenges in fully supporting emerging database systems due to semantic or syntax discrepancies. This aligns with our observations that SQLancer's generated queries for QuestDB are limited to several simple query structures. Our approach leverages enhanced differential testing to expand the testing scope through clause mapping, thereby achieving a much higher number of unique query plans.

\begin{figure}[t]
    \centering
    \includegraphics[scale=0.44]{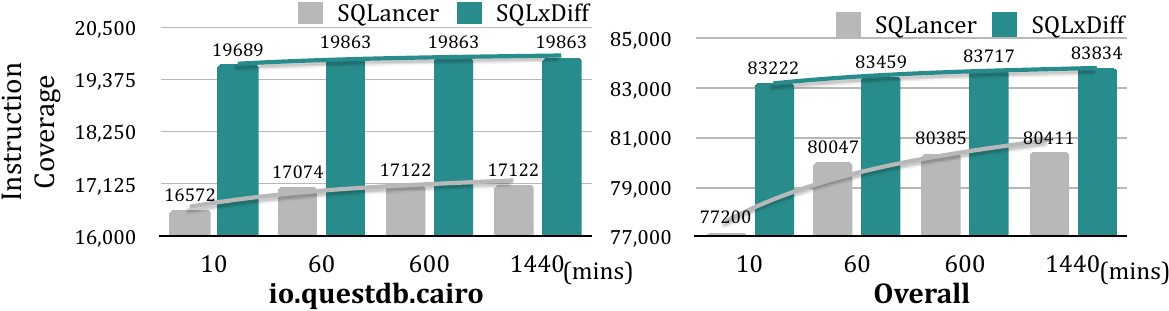}
    \caption{Code Coverage Comparison with SQLancer}
    \label{fig:jacoco}
\end{figure}

\paragraph{\textbf{Coverage}.} 
Code coverage is a widely used metric to evaluate the effectiveness of database system testing approaches~\cite{rigger2020finding, Rigger2020PQS}. We evaluate the code coverage of \name and SQLancer on QuestDB over 24 hours (as suggested in previous work~\cite{klees2018evaluating}) using the publicly available JaCoCo~\cite{jacoco} library. We present two statistics: overall code coverage and relevant code coverage focused on the SQL engine (\textie \textit{io.questdb.cairo} class) of QuestDB. The focused data excludes components not involved in query processing, such as functionality for establishing connections and authentications. Figure~\ref{fig:jacoco} shows that \name surpasses SQLancer in either overall code coverage or focused code coverage. In other classes, \name also achieves comparable code coverage to SQLancer. There have been some concerns about using code coverage as a convincing metric for evaluating testing approaches as stated in a previous study~\cite{inozemtseva2014coverage}. Higher coverage does not necessarily detect more bugs without a proper test oracle. 

\section{Related Work}

We briefly summarize the most relevant related work.



\paragraph{Detecting logic bugs in database systems.} Logic bugs, which refer to incorrect results returned by database systems, are difficult to detect as they require a \emph{test oracle}, a mechanism that decides whether the test case's result is expected. Song et al. proposed the oracle, \emph{Differential Query Execution} (DQE)~\cite{song2023testing}, to detect logic bugs in database systems by checking whether SQL queries with the same predicate access the same rows.
Rigger et al. proposed the oracles 
\emph{Pivoted Query Synthesis} (PQS)~\cite{Rigger2020PQS}, \emph{Non-Optimizing Reference Engine Construction} (NoREC)~\cite{Rigger2020NoREC}, and TLP~\cite{rigger2020finding} to detect logic bugs in relational database systems by checking results' consistency of several related queries, and have found hundreds of bugs. To generate test cases for such test oracles, SQLRight~\cite{liang:sqlright} leverages code coverage to guide the test case generation, and \emph{Query Plan Guidance} (QPG)~\cite{ba2023testing} guides the test cases toward unseen query plans aiming to generate the test cases that trigger diverse behaviors. 
Rather than finding bugs in relational database systems, we focus on emerging databases which have differences from each other and also with relational databases. Thus, the challenge is how to have effective bug finding including logic bugs given the differences between the database systems.


\paragraph{Differential testing.} 
Another line of research to detect logic bugs is differential testing~\cite{mckeeman1998differential}, which detects bugs by executing a test case using multiple versions or instances of systems that implement the same semantics, and any discrepancy indicates a potential bug in one of these systems. Researchers have utilized this method to detect bugs across various domains, such as web services~\cite{chapman2011automated}, Java Virtual Machine (JVM) implementations~\cite{chen2016coverage}, compilers~\cite{yang2011finding}, and network protocols~\cite{brumley2007towards}. Differential testing was applied to testing database systems as a system called \emph{RAGS}~\cite{slutz1998massive}, which executes a query on multiple different database systems and compares their results. \emph{APOLLO}~\cite{jung:apollo} also applies differential testing to detect performance issues by comparing the execution time of the same query on different versions of the same database system. In this work, we propose a novel approach to enhance the scalability of differential testing, making it applicable for uncovering logic bugs in emerging database systems that may have nontrivial dissimilarities from traditional relational databases.



\paragraph{Detecting memory errors in database systems.} Most previous methods for testing database systems have focused on memory errors, which does not require an explicit test oracle. Grey-box fuzzers, such as AFL~\cite{afl}, use mutation and code coverage as the fuzzing strategy. However, for database systems, the mutation methods typically incur invalid test cases due to the constraints of SQL grammar. 
Squirrel~\cite{zhong2020squirrel} uses a syntax-preserving mutation method to increase the rate of valid test cases during mutation. Generation-based methods, such as SQLSmith~\cite{sqlsmith}, DynSQL~\cite{jiang2023dynsql}, and ADUSA~\cite{liu2022automatic}, generate test cases according to a grammar. 
Griffin~\cite{fu2022griffin} uses a grammar-free test case generation method to alleviate the human effort to construct grammar for database systems. While these works have generated test cases efficiently, they have not tackled the test oracle problem to find logic bugs.

\section{Conclusion}

Testing emerging database systems to find logic and other bugs is challenging even when restricted to SQL-like systems as each has differences in syntax and semantics. The core insight behind our work is recognizing that emerging database systems can be viewed as extensions of relational database systems, sharing substantial functionality. We propose an extension of differential testing using relational database systems as the ground truth and show how to deal with the challenges with our approach combining finding shared clauses, expanding clauses with mappings, and a simple query generator that has broad applicability with shared and mapped clauses. We applied \name to four popular emerging database systems, uncovering a total of \totalbug unknown bugs, including \logicbug logic bugs and \errorbug internal errors. Vendors have confirmed or fixed 55 of these bugs. We believe that \name shows the feasibility of a unified testing framework for emerging database systems, which is important for enhancing their robustness and reliability.


\clearpage
\balance
\bibliographystyle{ACM-Reference-Format}
\bibliography{sample}
\appendix

\end{document}